\documentclass[
    ,final            
  ]
  {aipproc}

\layoutstyle{6x9}

\begin{document}
\title{Astrophysical Implications of Turbulent Reconnection: from cosmic
rays to star formation}


\author{A. Lazarian}{address={Department of Astronomy, University of Wisconsin, 475 N. Charter
St., Madison, WI 53706; lazarian@astro.wisc.edu}}

\begin{abstract}
Turbulent reconnection allows fast magnetic reconnection of
astrophysical magnetic fields. This entails numerous astrophysical
implications and opens new ways to approach long standing problems.
I briefly discuss a model of turbulent reconnection within which 
the stochasticity of 3D magnetic
field enables rapid reconnection through both allowing multiple reconnection
events to take place simultaneously and by restricting the extension of
current sheets. In fully ionized gas the model in Lazarian
\& Vishniac 99 predicts reconnection rates
that depend only on the intensity of turbulence. In partially ionized gas
a modification of the original model in 
Lazarian, Vishniac \& Cho 04 predicts 
the reconnection rates that, apart from the turbulence intensity 
depend on the degree of ionization. 
In both cases the reconnection
may be slow and fast depending on the level of turbulence in the system.
As the result, the reconnection gets bursty, which provides a possible
explanation to Solar flares and possibly to gamma ray busts. 
The implications of the turbulent reconnection model 
have not been yet studied in sufficient detail.
I discuss first order Fermi acceleration
of cosmic ray that takes place as the oppositely directed magnetic fluxes
move together. This acceleration would work in conjunction with the 
second order Fermi acceleration that is caused by turbulence in the 
reconnection
region. In partially ionized gas the stochastic reconnection enables
 fast removal of magnetic flux from star forming molecular clouds.

\end{abstract}
\maketitle

\section{What is magnetic reconnection?}

Magnetic fields play key role for many Astrophysical processes like
star formation, cosmic ray transport and acceleration, accretion etc.
As a rule, magnetic diffusivity is slow over huge astrophysical scales
and therefore frozenness of magnetic field provides an excellent
approximation (see Moffatt 1978). Magnetic field lines in astrophysical
highly conducting fluids
act as elastic
threads that are moved together with the fluid.
However, fluid motions are likely
to create knots in which magnetic threads will be
pressing against each other. If the state
of frozenness is not violated at these knots
this would result in the
formation of felt or Jello-like 
structure of magnetic field that is favored by some
astrophysicists (Cox 2004). If, however, crossing
magnetic field lines can change their topology, the dynamics of magnetized
fluid is {\it completely} different. 

The situation is rather dramatic. In fact, we cannot claim that we 
understand the
dynamics
of magnetized astrophysical fluids if we cannot figure out what picture 
is correct, i.e. whether astrophysical fluids act as a  magnetic Jello
or have fluid type behavior (see more below). 
Note, that adopting the first alternative would
also mean kiss of death to the contemporary dynamo theories,
i.e. to the theories that  explain magnetism of different 
astrophysical
objects from stars to galaxies by appealing to the amplification of 
magnetic fields by fluid motions (see Parker 1979). 

A naive answer to the question of whether astrophysical magnetic field 
can change the topology is positive. Indeed, as oppositely directed 
magnetic fields come 
together
the diffusion of magnetic field gets important inducing changes of 
magnetic
topology. The issue is the rates at which this topology can change. 
And this is
the key issue around which many decades of controversy about fast reconnection
are centered (see Biskamp 2000, Priest and Forbes  2000).

\begin{figure}
\includegraphics[width=.5\textwidth]{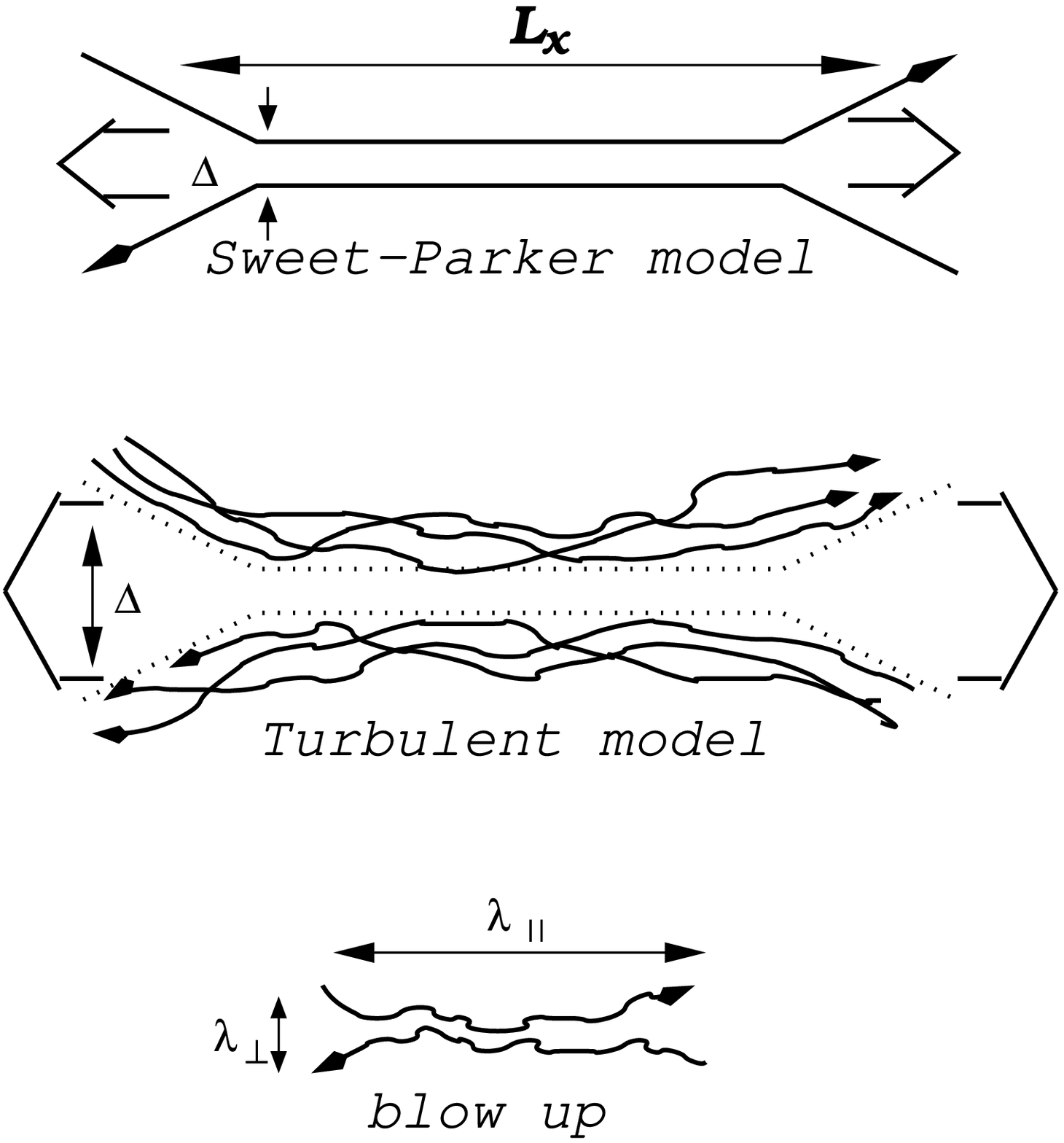}
\includegraphics[width=0.5\textwidth]{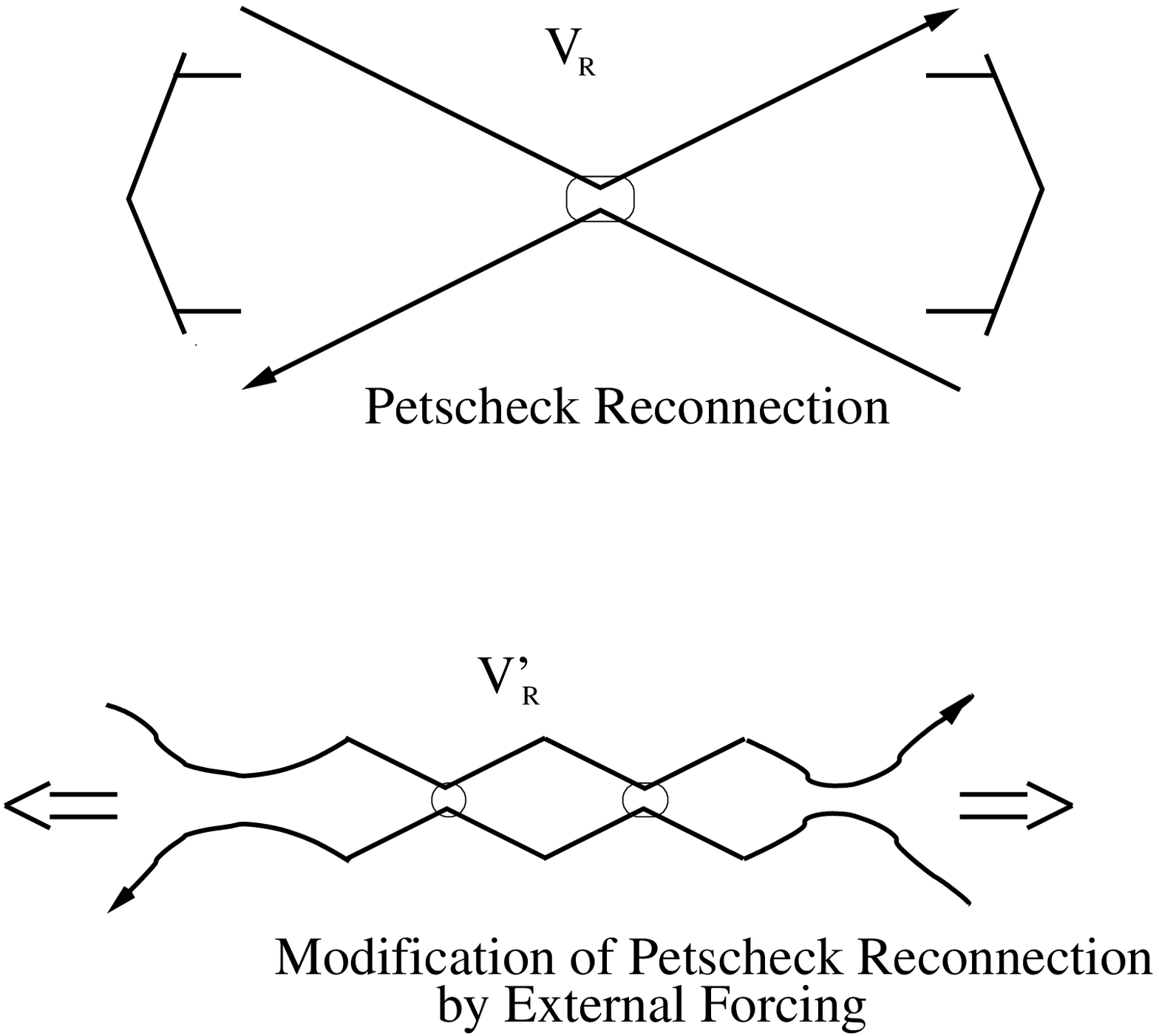}
\caption{ {\bf Left panel} {\it Upper plot}: 
Sweet-Parker model of reconnection. The outflow
is limited by a thin slot $\Delta$, which is determined by Ohmic 
diffusivity. The other scale is an astrophysical scale $L\gg \Delta$.
{\it Middle plot}: Turbulent reconnection model that accounts for the 
stochasticity
of magnetic field lines. The outflow is limited by the diffusion of
magnetic field lines, which depends on field line stochasticity.
{\it Low plot}: An individual small scale reconnection region. The
reconnection over small patches of magnetic field determines the local
reconnection rate. The global reconnection rate is substantially larger
as many independent patches come together. {\bf Right panel} {\it Upper plot}:
Petscheck reconnection scheme has a magnetic diffusion region (rectangular
area at the tip of the magnetic field line bending) for which both
the  longitudinal and transversal dimensions are determined 
by the Ohmic diffusivity.
To enable $V_R\sim V_A$ the model requires field line opening over the whole astrophysical scale
involved, which is difficult to satisfy in practice. {\it Lower plot}:
External forcing, e.g. the forcing present in the ISM, is likely
to close the opening required by the Petscheck model. In this case
the global outflow constraint is not satisfied and the resulting
reconnection speed is $V_R'\ll V_A$. As the result the Petscheck reconnection
cannot operate steadily and therefore cannot deal with the amount of flux
that, for instance, an astrophysical dynamo would require to reconnect.}
\end{figure}

The most natural and robust
scheme which was suggested first in the literature is the 
well known
Sweet-Parker model of reconnection (Parker 1957, Sweet 1958). 
Illustrated in Figure 1 this model of reconnection has two
oppositely directed magnetic 
fluxes
that get into contact over a scale $L$. While for most of
the volume the magnetic field lines are frozen in, the magnetic diffusivity 
is
important within the current sheet of thickness $\Delta$ (see Fig.~1).  
The {\it local} velocity at 
which the change of
magnetic topology takes place, i.e. the reconnection velocity $V_R$, 
is given
by
\begin{equation}
V_R=\eta/\Delta,
\end{equation}
where $\eta$ is the resistivity coefficient\footnote{The coefficient of 
magnetic
field diffusivity in a fully ionized plasma is 
$\eta=c^2/(4\pi \sigma)\approx 10^{13}T^{-3/2}$~cm$^2$ s$^{-1}$, where $\sigma\approx
10^7 T^{3/2}$ s$^{-1}$ is the plasma
conductivity and $T$ is electron temperature measured in Kelvins.
The characteristic time
for field diffusion through a plasma slab of size $y$ is
$y^2/\eta$, which is large for any ``astrophysical'' $y$. }. One may notice 
that the local reconnection
velocity can be arbitrary large provided that  $\Delta$ is sufficiently 
small. However,
if we are interested in reconnecting sufficient amounts of magnetic 
flux we should
consider steady state reconnection. Such a reconnection requires that the
fluid
and shared flux\footnote{It is easy to generalize the idealised 2D picture
above to three dimensions. In 3D the current flows 
along the shared magnetic flux. This flux is being ejected together with 
the fluid from the diffusion region.}
be removed from the reconnection region in order to allow fresh oppositely 
directed
magnetic field lines to get into close contact. This imposes 
the {\it global} constraint
upon the reconnection rate :
\begin{equation}
V_R L=V_A \Delta,
\end{equation}
where the removal of fluid happens with the 
Alfv\'en velocity
$V_A$ and the thickness of the outflow is determined by $\Delta$. As the 
result of these two constraints acting simultaneously we
get the well-known Sweet-Parker reconnection rate:
\begin{equation}
V_R=V_A R_L^{-1/2},
\end{equation}
where $R_L\equiv (L V_A/\eta)$ is the so-called Lundquist number, which is an
analog of the hydrodynamic Reynolds number. The enormous scales of 
astrophysical
systems, i.e. large $L$ and relatively low values of magnetic diffusivity 
$\eta$ make
$R_L$ really huge for most of astrophysically interesting situations. 
Indeed, the 
corresponding $R_L$ may range from $10^{10}$ to $10^{20}$ for 
interstellar gas.
As the result the reconnection rates get so insignificant that one should 
conclude that
forgetting about reconnection {\it if it happens with the Sweet-Parker 
rate} is an excellent
astrophysical approximation.  This entails many interesting
conclusions. One of them is that 
all current magnetohydrodynamic
simulations are useless, as the present day computer codes  are 
diffusive and do not describe Jello-type fluids. On the contrary,
they induce efficient magnetic reconnection allowing magnetic
field  to change its topology on dynamical times. 

Is the situation that bad? How can one explain Solar
flares, which indicate that the real world astrophysical magnetic fields 
do change their topology over short time scales (see Dere 1996)?

The problem of the Sweet-Parker reconnection scheme is a huge disparity of
scales as $L$ is an astrophysical scale, while $\Delta$ is given by 
microphysics.
The attempts to remove this disparity resulted in Petscheck (1964) scheme of
reconnection (see Fig.~1, left panel). 
In this model the both scales involved are determined by
microphysics. The reconnection happens over tiny portion of the length of
magnetic field lines, while the rest of the magnetic flux forms an X-type
configuration that allows an easy escape of the reconnected flux and fluid.

Formally, the Sweet-Parker scheme and the Petscheck scheme are of the
same class of the reconnection models in which the local and global 
reconnection
rate must coincide. This imposes stringent constraints on the geometry of the
Petscheck model. Indeed, the inevitable requirement of Petscheck 
scheme is that 
X-type structure must be sustainable over the scale at which opposite 
fluxes get together. 
If external forcing decreases X-type opening, the global outflow conditions
cannot be satisfied and the reconnection is chocked. Surely, the
reconnection can happen fast over short time scales, but for reconnecting
large amounts of flux a quasi-steady fast reconnection is required. 
This means
that, for instance, interstellar magnetic fields should be configured to 
form X-points
over many parsecs to enable fast reconnection.

Petscheck scheme has
been surrounded by many controversies from the time that it originated. 
It is clear
at this moment that it cannot be sustained  at large $R_L$ 
for smooth resistivities.
Whether or not anomalous effects, e.g. those related to Hall term, 
can provide reconnection at rates comparable with the Alfv\'en speed
is hotly debated\footnote{ For instance, the necessary condition
for the anomalous effects to be important, e.g. for
that the electron mean free path is less than the current sheet thickness (see 
Trintchouk et al. 2003) is difficult to satisfy for the ISM, where
the Sweet-Parker current sheet thickness is typically {\it much} larger
than the ion Larmor radius.} (see Biskamp, Schwarz \& Drake 1997, 
Shay et al. 1998, Shay \& Drake 1998,
Bhattacharjee, Ma \& Wang 2001). We feel, however, that 
the issue of
satisfying boundary outflow conditions is  the most controversial element
in applying Petscheck scheme to astrophysical conditions. If very special
global geometry or magnetic fluxes, e.g. convex magnetized regions, 
is required to enable reconnection, then
for a generic astrophysical case the reconnection is slow.  

\section{How can turbulence enhance reconnection?}

Reconnection involves dissipation and diffusion. Therefore it is natural to
wonder whether turbulence can enhance the reconnection rate. 
The notion that magnetic field stochasticity might affect 
current sheet structures is not unprecedented.  In earlier
work Speiser (1970) showed that in collisionless plasmas the
electron collision time should be replaced with the time a
typical electron is retained in the current sheet.  Also
Jacobson \& Moses (1984) proposed that current diffusivity should be modified
to include diffusion of electrons across the mean field due
to small scale stochasticity.  These effects will usually be small
compared to effect of a broad outflow zone containing both
plasma and ejected shared magnetic flux.  Moreover,
while both of these effects
will affect reconnection rates\footnote{It looks that current experiments
to study reconnection do measure such enhancements related to magnetic
field stochasticity (Ji 2005).}, they are not sufficient to 
produce reconnection speeds comparable to the Alfv\'en speed
in most astrophysical environments.  

"Hyper-resistivity" discussed in a number of works (e.g. Strauss 1985, 
Bhattacharjee \& Hameiri 1986, Hameiri \& Bhattacharjee 1987)
 is a more subtle attempt to derive fast reconnection
from turbulence within the context of mean-field resistive MHD.  
The form of the parallel
electric field can be derived from magnetic helicity conservation.  Integrating
by parts one obtains a term which looks like an effective resistivity
proportional to the magnetic helicity current.  There are several assumptions
implicit in this derivation, but the most important problem is that by
adopting a mean-field approximation one is already assuming some sort of
small-scale smearing effect, equivalent to fast reconnection.  Strauss (1988)
partially circumvented this problem by examining the effect of tearing
mode instabilities within current sheets.  However, the resulting reconnection
speed enhancement is roughly what one would expect based simply on the
broadening of the current sheets due to internal mixing.  This effect
does not allow us to evade the constraints on the global
plasma flow that lead to slow reconnection speeds, a point which
has been demonstrated numerically (Matthaeus \& Lampkin 1985)
and analytically (LV99). 

A different approach has been adopted in Lazarian \& Vishniac (1999, 
henceforth LV99). The starting point there
is that turbulence is a generic state of astrophysical fluids (see
Armstrong, Rickett, Spangler 1995, Lazarian \& Pogosyan
2004, Lazarian 2004). 
MHD turbulence guarantees the presence of a stochastic field component,
although its amplitude and structure clearly depends on the model we adopt
for MHD turbulence (see review by Cho, Lazarian \& Vishniac 2003a and
references therein), as well as the specific environment of the field. 
We consider the case in which there exists a large scale,
well-ordered magnetic field, of the kind that is normally used as
a starting point for discussions of reconnection.  
In addition, we expect that the field has some small scale `wandering' of
the field lines.  On any given scale the typical angle by which field
lines differ from their neighbors is $\phi\ll1$, and this angle persists
for a distance along the field lines $\lambda_{\|}$ with
a correlation distance $\lambda_{\perp}$ across field lines (see Fig.~1).

The modification of the global constraint induced by mass conservation
 in the presence of
a stochastic magnetic field component 
is self-evident. Instead of being squeezed from a layer whose
width is determined by Ohmic diffusion, the plasma may diffuse 
through a much broader layer, $L_y\sim \langle y^2\rangle^{1/2}$ (see Fig.~1),
determined by the diffusion of magnetic field lines.  This suggests
an upper limit on the reconnection speed of 
$\sim V_A (\langle y^2\rangle^{1/2}/L_x)$. 
This will be the actual speed of reconnection if
the progress of reconnection in the current sheet does not
impose a smaller limit. The value of
$\langle y^2\rangle^{1/2}$ can be determined once a particular model
of turbulence is adopted, but it is obvious from the very beginning
that this value is determined by field wandering rather than Ohmic
diffusion as in the Sweet-Parker case.

What about limits on the speed of reconnection that arise from
considering the structure of the current sheet?
In the presence of a stochastic field component, magnetic reconnection
dissipates field lines not over their  entire length $\sim L_x$ but only over
a scale $\lambda_{\|}\ll L_x$ (see Fig.~1), which
is the scale over which magnetic field line deviates from its original
direction by the thickness of the Ohmic diffusion layer $\lambda_{\perp}^{-1}
\approx \eta/V_{rec, local}$. If the angle $\phi$ of field deviation
does not depend on the scale, the local
reconnection velocity would be $\sim V_A \phi$ and would not depend
on resistivity. In LV99 we claimed that $\phi$ does depend on scale. 
Therefore the {\it local} 
reconnection rate $V_{rec, local}$ is given by the usual Sweet-Parker formula
but with $\lambda_{\|}$ instead of $L_x$, i.e. $V_{rec, local}\approx V_A 
(V_A\lambda_{\|}/\eta)^{-1/2}$.
It is obvious from Fig.~1 that $\sim L_x/\lambda_{\|}$ magnetic field 
lines will undergo reconnection simultaneously (compared to a one by one
line reconnection process for
the Sweet-Parker scheme). Therefore the overall reconnection rate
may be as large as
$V_{rec, global}\approx V_A (L_x/\lambda_{\|})(V_A\lambda_{\|}/\eta)^{-1/2}$.
Whether or not this limit is important depends on
the value of $\lambda_{\|}$.  

The relevant values of $\lambda_{\|}$ and $\langle y^2\rangle^{1/2}$ 
depend on the magnetic field statistics. This
calculation was performed in LV99 using the Goldreich-Sridhar (1995, 
henceforth) model
of MHD turbulence, which has been supported by numerical simulations
(see review by Cho, Lazarian \& Vishniac 2003z and references therein).
For instance, for the GS95 spectrum, which
according to Cho \& Lazarian (2003) persists for Alfv'enic mode
even in compressible
MHD turbulence with a large-scale high-amplitude driving,
the upper limit on the reconnection speed was 
\begin{equation}
V_{r, up}=V_A \min\left[\left({L_x\over l}\right)^{\frac{1}{2}}
\left({l\over L_x}\right)^{\frac{1}{2}}\right]
\left({v_l\over V_A}\right)^{2},
\label{main}
\end{equation}
where $l$ and $v_l$ are the energy injection scale and
turbulent velocity at this scale respectively.
In LV99 other processes that can impede
reconnection were found to be less restrictive. For
instance, the tangle of reconnection field lines crossing the
current sheet will need to reconnect repeatedly before individual
flux elements can leave the current sheet behind.  The rate at which
this occurs can be estimated by assuming that it constitutes the
real bottleneck in reconnection events, and then analyzing each
flux element reconnection as part of a self-similar system of
such events.  This turns out to impede the reconnection.  
As the result LV99 
concludes that (\ref{main}) is not only an
upper limit, but is the best estimate of the speed of reconnection.

We stress that the enhanced reconnection efficiency in turbulent
fluids is only present if 3D reconnection is considered. In
this case Ohmic diffusivity fails to constrain the reconnection process 
as many
field lines simultaneously enter the reconnection region. The
number of lines that can do this increases with the decrease of
resistivity and this increase overcomes the slow rates of
reconnection of individual field lines. It is impossible to achieve
a similar enhancement in 2D (see Zweibel 1998) since field lines
can not cross each other.

It worth mentioning that in a recent paper Kim \& Diamond (2001) 
addressed the problem of stochastic
reconnection by calculating the turbulent diffusion rate for magnetic flux
inside a current sheet.  They obtained similar turbulent 
diffusion rates for both two dimensional and reduced three 
dimensional MHD. In both cases the presence of turbulence had a
negligible effect on the flux transport.  The authors pointed out
that this would prevent the anomalous transport of magnetic flux 
within the current
sheet and concluded that both 2D and 3D stochastic reconnection proceed
at the Sweet-Parker rate even if individual small scale reconnection
events happen quickly.

However turbulent diffusion rates within the current sheet are irrelevant 
for the process of stochastic reconnection (see discussion in Lazarian,
Vishniac \& Cho 2004, henceforth LVC04). The basic claim in LV99 
is that realistic magnetic field topologies allow multiple
connections between the current sheet and the exterior environment, which would
persist even if the stochastic magnetic field lines were stationary 
("frozen in 
time") before reconnection.  This leads to global outflow constraints which
are weak and do not depend on the properties of the current sheet.  
In particular,
the analysis in LV99 assumed that the current sheet thickness is
determined purely by ohmic dissipation and that turbulent diffusion of the 
magnetic
field is negligible inside, and outside, the current sheet. 

\section{Does partial ionization matter?}

The notion that the reconnection may be  different in fully ionized and
partially ionized gas is not new (see Parker 1977).
Reconnection in partially ionized gases has been addressed by
various authors (see Naidu, McKenzie \& Axford 1992, Zweibel \& Brandenburg
1997).  In a more
recent study Vishniac \& Lazarian (1999)
considered the diffusion of neutrals away from the reconnection zone
assuming anti-parallel magnetic field lines (see also Heitsch \& Zweibel
2003a).
The ambipolar reconnection rates obtained by Vishanic \& Lazarian (1999a), 
although large
compared with the Sweet-Parker model, are insufficient either
for fast dynamo models or for the ejection of magnetic flux prior
to star formation.  In fact, the increase in the reconnection speed
stemmed entirely from the
compression of ions in the current sheet, with the consequent enhancement of
both recombination and ohmic dissipation.  This effect is small
unless the reconnecting magnetic field lines are almost exactly
anti-parallel (Vishniac \& Lazarian 1999a, LV99, Heitsch \& Zweibel 2003b). 
 Any dynamically significant shared field component
will prevent noticeable plasma compression in the current
sheet, and lead to speeds practically indistinguishable from the standard
Sweet-Parker result.  Since generic reconnection regions will have
a shared field component of the same order as the reversing component,
ambipolar diffusion does not
change reconnection speed estimates significantly.

What is the effect of neutrals on stochastic reconnection?
It is obvious that neutrals should modify the turbulent cascade at 
sufficiently small scales. Indeed, while ions move with magnetic
fields, neutrals in partially ionized plasma are moved via
collisions with ions. The incomplete coupling of ions and neutrals creates
friction that damps MHD turbulence. According to LVC04
 this gives rise to an interesting new regime of turbulence
that creates magnetic stochasticity at scales much smaller
than the scale at which neutrals damp turbulent
kinetic energy cascade. Eventually,
according to LVC04, at very small scales
ions decouple from neutrals completely and produce bursts of intermittent
in space and time MHD turbulence that involves only ions\footnote{This
can explain observations by Spangler (1991, 1999) that fail to detect
the existence of cut-off of magnetic perturbations in a partially ionized
interstellar gas.}. While the latter
conclusion is still requires testing with two fluid code, the other
conclusions of the model, e.g. the formation of magnetic fluctuations at
scales smaller than the scale of viscous damping, the spectra of 
magnetic field  and kinetic energy at small scales, the
intermittency of magnetic fluctuations, have been successfully tested
numerically (Cho, Lazarian \& Vishniac 2002, 2003b). With this encouragement
from numerics LVC04 applied the expected scaling of magnetic fluctuations
to the LV99 stochastic reconnection model and obtained the reconnection
rates that depend not only on the intensity of MHD turbulence, but also
on the parameters of the partially ionized gas, e.g. 
on ionization ratio, ion-neutral collision time. Although more restrictive
than those given by eq.(\ref{main}) the rates calculated for different
idealized
phases of interstellar medium, namely, for Warm Ionized Medium, Warm Neutral
Medium, Cold Neutral Medium, Molecular Cloud and Dark Cloud are fast enough
to enable rapid changes of magnetic field topology.      

\section{How can turbulent reconnection account for Solar flares?}

Solar observations indicate that reconnection can be both slow and fast.
Solar flares, which are generally believed to be powered by magnetic
reconnection, can happen if oppositely directed magnetic flux can be
accumulated for a while without being reconnected. The dynamics
of solar flaring
seem to show that reconnection events start from some limited
volume and spread as though
a chain reaction from the initial reconnection region induced
a dramatic change in the magnetic field properties. Indeed, 
solar flaring happens as if the
resistivity of plasma were increasing dramatically as plasma
turbulence grows (see Dere 1996 and references therein).
According to LV99 this is a consequence of
the increased stochasticity of the field lines rather than
any change in the local resistivity.

If we start with nearly laminar magnetic fields the flaring of 
reconnection is a natural consequence. 
Indeed, when turbulence is negligible, i.e. $v_l\rightarrow 0$, the
field line wandering is limited to the Sweet-Parker current sheet
and the Sweet-Parker reconnection scheme takes over.
However, the release of energy due to reconnection is likely
to induce turbulence in astrophysical fluids. If this is
the case, the level of turbulence grows and the turbulent
reconnection takes over (LV99). In fact, one may say that 
{\it reconnection instability} takes place as the higher
the rate of energy release due to reconnection, the faster
reconnection is, which in its turn increases the rate
of energy release.  It is possible to
show that this positive feedback induces finite time instability
(Vishniac \& Lazarian 1999b), which
agrees well with the observations of Solar flares.

\section{How can turbulent reconnection accelerate cosmic rays?}

There are several ways how magnetic 
reconnection can accelerate cosmic rays. It is well known
that electric fields in the current sheet can do the job. For Sweet-Parker
reconnection this may be an important process in those exceptional
instances when Sweet-Parker reconnection is fast in
astrophysical settings. For Petscheck reconnection only an insubstantial
part of magnetic energy is being released within the reconnection
zone, while bulk of the energy is being released in shocks that
support  X-point. Therefore one would expect the shock acceleration
of cosmic rays to accompany Petscheck reconnection. 

\begin{figure}
\includegraphics[width=.7\textwidth]{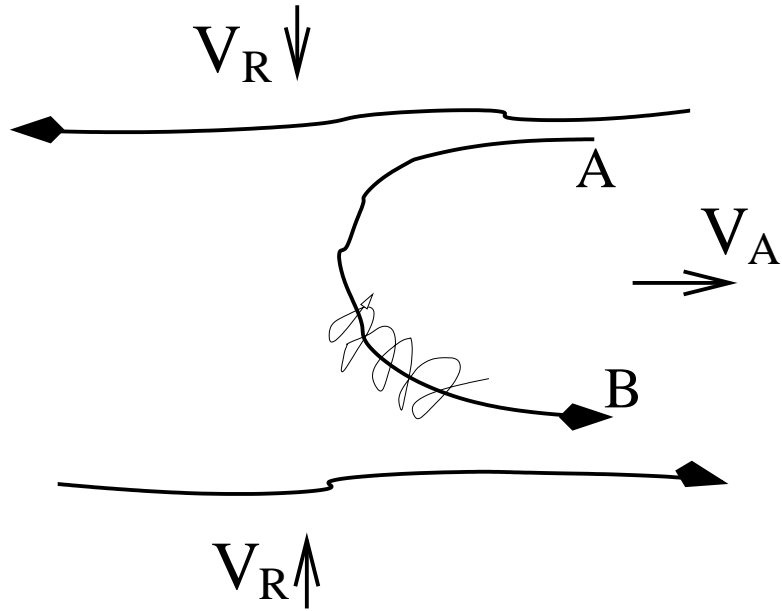}
\caption{Cosmic rays spiral about a reconnected magnetic
field line and bounce back at points A and B. The reconnected
regions move towards each other with the reconnection velocity
$V_R$. The advection of cosmic rays entrained on magnetic field
lines happens at the outflow velocity, which is in most cases
of the order of $V_A$. Bouncing at points A and B happens
because either of streaming instability or turbulence in the
reconnection region.}
\end{figure}

Similarly to the Petscheck scheme 
the turbulent reconnection process assumes that only
small segments of magnetic field lines enter the reconnection zone
and are subjected to Ohmic annihilation. Thus only small fraction of
magnetic energy, proportional to ${R}_L^{-2/5}$ (LV99), is
released in the current sheets. The rest of the energy is
released in the form of non-linear Alfv\'en
waves that are generated as reconnected magnetic field lines straighten
up. Such waves are likely to cause second order Fermi acceleration.
This idea was briefly discussed in Lazarian et al. (2001) in relation
to particle acceleration during the gamma-ray burst events. In addition,
large amplitude Alfv\'enic motions in low $\beta$, i.e. magnetically
dominated, plasmas are likely to induce shocks (see Beresnyak, Lazarian
\& Cho 2005), which can also cause particle acceleration.

However, the most interesting process is the first-order Fermi acceleration
that is intrinsic to the turbulent reconnection. To understand it
consider a particle entrained on a reconnected 
magnetic field line (see Fig.~2). This particle
may bounce back and force between magnetic mirrors formed by oppositely
directed magnetic fluxes moving towards each other with the velocity
$V_R$. Each of such bouncing will increase the energy of a particle
in a way consistent with the requirements of the first-order Fermi
process. The interesting property of this mechanism that potentially
can be used  to test observationally the idea is that the 
resulting spectrum is different from those arising from shocks.
Gouveia Dal Pino \& Lazarian (2003) used particle acceleration
within turbulent reconnection regions to explain
the synchrotron power-law spectrum arising from the flares of the
microquasar GRS 1915+105. Note, that the mechanism acts in the
Sweet-Parker scheme as well as in the scheme of turbulent reconnection.
However, in the former the rates of reconnection and therefore the
efficiency of acceleration are marginal in most cases.

As a note of caution we would like to warn our reader that the detailed
theory of cosmic ray acceleration during turbulent reconnection that 
accounts for both first and second order Fermi acceleration is yet to
be developed\footnote{The following points that must be taken into account 
while calculating cosmic ray acceleration. First of all, the streaming
instability is partially suppressed due to a non-linear interaction with
ambient MHD turbulence (Yan \& Lazarian 2002, 2004, Farmer \& Goldreich 2004).
For cosmic ray scattering from turbulent fluctuations it is absolutely
essential that the proper
scaling of MHD modes is taken (Chandran 2000, Yan \& Lazarian 2002).}
However, this seems to be a interesting way for approaching
the cosmic ray acceleration problem.

\section{What are other implications?}

{\bf Star formation}

It is well known that the frozenness of
magnetic field condition should be violated during star formation.
Otherwise, stars would have magnetic fields orders of magnitude
larger than the observed values. Magnetic field removal is also
an important ingredient of the evolution of star-forming 
magnetized clouds. While usually these
issues are addressed through appealing to ambipolar
diffusion (see Tassis \& Mouschovias 2005), 
turbulent reconnection in partially ionized gas
described in LVC04 is well suited for doing the job. Note, that
ambipolar diffusion is too slow at least for supercritical
star formation that takes place over free fall time. This problem
does not arise for turbulent reconnection. While the very idea that
reconnection can be an important component of star formation was
advocated for a while by Frank Shu, the quantitative implementation
of it was impeded by the absence of reliable estimates of the
reconnection rate.

{\bf Transport processes}

Various astrophysical transport processes are affected by reconnection.
Let us start with the process of {\it heat transport}. Cho et al. (2003)
has shown numerically 
that heat can be efficiently transported by turbulent eddies
within magnetized fluid thus preventing cooling flows from forming in
clusters of galaxies. Such a mixing is difficult to visualize unless
magnetic fields change their topology rather than form magnetic knots
(Lazarian \& Cho 2004). Similarly {\it transport of matter} is affected
by reconnection. Without it cosmic rays, ions, charged dust particles
would always be frozen in on ``their own'' magnetic field lines,
which would drastically impede mixing processes
on astrophysical scales. In addition,
{\it transport of angular momentum} in accretion disks 
is expected to be very different depending on the rate of reconnection.
Fast turbulent reconnection should prevent direct magnetic connection
of elements that are at very different radial distances from the center.
This does limit the efficiency of the angular momentum transport.

{\bf Dynamo}

Mean field dynamo is an elegant theory explaining the origin of astrophysical
magnetic fields via
amplification of the seed field by fluid motions (see Mofatt 1978, 
Parker 1979). 
The original seed
field may be tiny and in some sense its value is frequently irrelevant provided
that an efficient dynamo has enough time to operate. Fast reconnection is
a necessary part the theory. Problems and approaches to mean field dynamo
are discussed in a recent review by Vishniac, Lazarian \& Cho (2003).
It  worth noting that the
stochastic reconnection dissipates only insignificant part of the magnetic
energy through Ohmic heating and therefore it does not remove the
constraint given by the helicity conservation. A mean field dynamo
model proposed by Vishniac \& Cho
(2002) takes explicitely this constraint into account.

Other implications, e.g. self-consistency of strong Alfv\'enic turbulence,
heating of electrons in accretion flows, and heating of interestellar
gas are discussed in LV99 and Lazarian
\& Vishniac (2000).

\section{How to test ideas above?}

It has been shown in the paper above that turbulent reconnection affects
many essential astrophysical processes. Therefore it is important to
test the model.

Implications of turbulent reconnection provide an indirect way
to test the model. Thus a way of testing the model is through
making quantitative predictions of the reconnection effects that can be 
tested. Solar flares, spectra of particles emerging during reconnection
are examples of such effects. A search for singnatures of Petscheck
reconnection, e.g. for the large scale X-points that should be observable,
provides another indirect way of searching for the truth.

Turbulent reconnection can be tested directly in laboratory. The problem
there is that the microscale and macroscale are usually not so far apart for
laboratory experiments. To distinguish between plasma turbulence
effects and MHD effects that our model deals with, the amplitudes
of magnetic perturbations should be substantial.

Numerical testing of turbulent reconnection looks at the moment as the most
promising way to go. With computers getting more powerful it is
feasible to test the predictions 
obtained in LV99. Testing reconnection rates in partially ionized gas
(LVC04) looks very challenging, but doable in future using two fluid
codes.

{\bf ACKNOWLEDGMENTS}\\
I am particularly grateful to Ethan Vishniac for great time of
working together on the problem of magnetic reconnection.
I acknowledge the support of the NSF grant AST 0307869 and
the Center for Magnetic Self-Organization in Laboratory and 
Astrophysical Plasmas.

\end{document}